\documentclass[aps,preprintnumbers,superscriptaddress,longbibliography,amsmath,amssymb,twocolumn,10pt,floatfix]{revtex4-2}

\usepackage{graphicx}
\usepackage{dcolumn}
\usepackage{bm}
\usepackage{xspace}
\usepackage{multirow}
\usepackage{booktabs}
\usepackage{xcolor}
\usepackage{xfrac}
\usepackage[caption=false]{subfig}

\usepackage[normalem]{ulem}

\graphicspath{{./plots/}}

\usepackage{hyperref}
\hypersetup{
    colorlinks=true,
    linkcolor=blue,
    filecolor=blue,      
    urlcolor=blue,
    pdftitle={Overleaf Example},
    pdfpagemode=FullScreen,
    }

\newcommand{\Xmax}{X_{\rm max}}

\newcommand{\Xrit}{X_{\rm RIT}}

\usepackage[mathlines]{lineno}

\usepackage{units}
\usepackage{comment}

\newcommand{\x}[1]{%
  {}$
  \kern-2\mathsurround 
  $
  \binoppenalty10000 \relpenalty10000 #1
  {}$
  \kern-2\mathsurround 
  $
}

\newcommand{\dummyfig}[1]{
  \fbox{
    \begin{minipage}[c][0.35\textwidth][c]{0.75\columnwidth}
      \centering{\url{#1}}
    \end{minipage}
  }
}

\newcommand{\dummyfigbi}[1]{
  \centering
  \fbox{
    \begin{minipage}[c][0.4\textwidth][c]{0.75\columnwidth}
      \centering{\url{#1}}
    \end{minipage}
  }
}

\newcommand{\dummyfigtri}[1]{
  \centering
  \fbox{
    \begin{minipage}[c][0.3\textwidth][c]{0.25\columnwidth}
      \centering{\url{#1}}
    \end{minipage}
  }
}

\begin{document}


\title{Average universal shower profile reconstruction using radio interferometry}
\date{\today}

\author{J.~Alvarez-Mu\~niz}
\affiliation{Instituto Galego de F\'\i{}sica de Altas Enerx\'\i{}as (IGFAE), Universidade de Santiago de Compostela, Santiago de Compostela, Spain}

\author{W.~R.~Carvalho~Jr.}
\affiliation{Faculty of Physics, University of Warsaw, Ludwika Pasteura 5, 02-093 Warsaw, Poland}

\author{R.~Concei\c{c}\~ao}
\affiliation{Departamento de F\'isica, Instituto Superior T\'{e}cnico, Universidade de Lisboa, Lisbon, Portugal}
\affiliation{LIP - Laborat\'orio de Instrumenta\c{c}\~ao e F\'isica Experimental de Part\'iculas, Lisbon, Portugal}

\author{D. Dias}
\affiliation{Departamento de F\'isica, Instituto Superior T\'{e}cnico, Universidade de Lisboa, Lisbon, Portugal}
\affiliation{LIP - Laborat\'orio de Instrumenta\c{c}\~ao e F\'isica Experimental de Part\'iculas, Lisbon, Portugal}

\begin{abstract}
Radio detection of extensive air showers enables near-continuous observation and precise measurements of the shower geometry and the depth of shower maximum, $X_{\rm max}$. Beyond $X_{\rm max}$, the longitudinal shower development follows a Universal Shower Profile (USP), whose shape parameters contain information on primary mass composition and hadronic interaction models. While previous studies have focused on event-by-event reconstruction of profile parameters, in this work we investigate the reconstruction of the \emph{average} USP using radio interferometric techniques. 

Using Monte Carlo simulations, we reconstruct the average longitudinal profile directly from radio data and extract the Gaisser--Hillas shape parameters $(R,L)$. We find that the radio-derived average profiles provide enhanced separation between primary masses and hadronic interaction models compared to that obtained from fluorescence-equivalent longitudinal profiles. These results demonstrate that radio interferometry can access higher-order information on the longitudinal shower development and that the use of the average USP significantly improves the sensitivity to composition and hadronic interaction studies at ultra-high energies.
\end{abstract}

\pacs{Valid PACS appear here}
\maketitle


\section{Introduction}
\label{sec:introduction}

The measurement of the longitudinal development of extensive air showers (EAS) provides access to the nature of ultra-high-energy cosmic rays (UHECR) and to the behavior of hadronic interactions at energies beyond those reachable in particle accelerators\,\cite{Engel:2011zzb}. The depth of shower maximum, $\Xmax$, has long been one of the primary observables used to infer mass composition and, in combination with ground-based observables, to test hadronic interaction models. Fluorescence measurements of $\Xmax$ have revealed an evolution of composition with energy~\cite{PierreAuger:2014gko,Kampert:2012mx}, and have exposed tensions with extrapolations to UHE of hadronic interaction models tuned to accelerator and collider data\,\cite{PierreAuger:2021qsd,PierreAuger:2024neu, Albrecht:2025kbb}.

Beyond $\Xmax$, the full longitudinal profile contains additional information. When the shower depth in the atmosphere $X$ is expressed relative to the depth of maximum ($X-\Xmax$) and the number of particles $N$ is normalised to the number of particles at maximum ($N/N_\mathrm{max}$), showers are known to follow an approximately universal shape known as the \textit{Universal Shower Profile} (USP)\,\cite{Andringa:2011zz,Andringa:2011ik}. The parameters governing the shape of the USP, obtained from Gaisser--Hillas function fits to the longitudinal profile, are sensitive to both the primary mass and the underlying hadronic interaction model\,\cite{Conceicao:2015toa}. In fluorescence experiments, this information is accessible, though statistically limited by the low duty cycle of optical detectors~\cite{PierreAuger:2018gfc}.

Radio detection offers a complementary approach with nearly continuous operation. Radio emission in EAS\,\cite{Huege:2016}, dominated by the geomagnetic\,\cite{Kahn:1966} and charge-excess mechanisms\,\cite{Askaryan:1962}, has enabled competitive angular resolution and precise $\Xmax$ reconstruction\,\cite{Buitink:2016nkf,Bezyazeekov:2018yjw,PierreAuger:2023lkx}. In addition, recent studies have explored the sensitivity of radio observables to features of the longitudinal profile\,\cite{DeHenau:2025zoh,deErrico:2026usf}. Radio interferometric techniques (RIT), which exploit signal coherence across antenna arrays, provide a direct way to reconstruct the shower development from radio measurements\,\cite{RIT1,RIT2}.

While the estimation of $\Xmax$ with radio measurements is well established\,\cite{Buitink:2016nkf,Bezyazeekov:2018yjw,PierreAuger:2023lkx,Corstanje:2025wbc}, the capability of radio detection techniques to robustly recover higher-order information encoded in the longitudinal profile remains under active investigation. Recent studies have demonstrated some sensitivity to individual shape parameters of the profile\,\cite{DeHenau:2025zoh,deErrico:2026usf}. Here, we extend this line of work by exploring the joint information contained in the two shape parameters $(L,R)$ of the USP, with $L$ determining the width of the shower profile and $R$ its skewness\,\cite{Andringa:2011zz}, enabling quantitative tests of primary-mass and hadronic-interaction models. A promising strategy is to consider the \emph{average} USP, suppressing event-by-event fluctuations and enhancing the sensitivity to systematic differences between primaries and models.

In this work, we show that the average USP can be reconstructed directly from radio data using radio interferometric techniques. Based on detailed Monte Carlo simulations, we compare radio-derived shape parameters of the USP longitudinal profiles with those of the electromagnetic shower development. We assess the sensitivity of these parameters to primary mass and hadronic interaction models when derived with RIT. Our results indicate that radio interferometry can access not only $\Xmax$ but also the shape of the average longitudinal profile, providing a high-duty-cycle complement to fluorescence measurements for UHECR composition determination and studies of hadronic interactions in EAS.

This paper is organized as follows. In Section~II, we describe the simulation framework and datasets used in this study. Section~III presents the methodology to reconstruct the longitudinal shower profile using information from the radio emission in EAS. In Section~IV, we construct the average USP. In Section~V, we discuss the sensitivity of the USP to primary cosmic-ray composition and hadronic interaction models. In Section~VI, we summarize the results and outline future prospects.

\section{Simulation framework and datasets}

To perform the proposed study, a Monte Carlo framework capable of simulating both the development of extensive air showers (EAS) and their associated radio emission is required. In particular, the simulation must model the propagation of the shower in the atmosphere, the production of radio emission, and the propagation of the resulting electric field to the observation level where it is detected by antennas. For this purpose we use \textsc{ZHAireS} (version 1.0.30a), which combines the \textsc{AIRES} (AIR-shower Extended Simulations) air-shower simulation package\,\cite{aires} for the tracking of the particles in the shower, with the radio emission algorithm based on the ZHS (Zas--Halzen--Stanev) formalism\,\cite{ZHS92,TimeDomainZHS,ZHAireS} for the calculation of the electric field emitted by a charged particle track. 

To explore the potential of radio interferometry to access the longitudinal development of EAS and to infer primary mass composition from the average USP, several simulated datasets were produced. The primary particle and hadronic interaction model were varied, considering proton and iron primaries, and SIBYLL~2.3d~\cite{sibyll} and QGSJET-II-04~\cite{qgsjet} as the high-energy interaction models. Four datasets were generated by combining the two primaries with the two hadronic interaction models, each containing $100$ simulated events. These datasets are referred to as \textit{SIB Proton}, \textit{QGS Proton}, \textit{SIB Iron}, and \textit{QGS Iron}.

The simulated air showers correspond to a primary energy of $10^{18}$~eV, arriving with a zenith angle $\theta=30^\circ$ and an azimuth angle $\phi=45^\circ$. A thinning procedure was employed, with a relative thinning level of $10^{-5}$, a weight factor of $0.06$, and an electromagnetic-to-hadronic weight ratio of $100$. The observation site is located in Malargüe, Argentina, at latitude $-35.20^\circ$, longitude $-69.20^\circ$, and at an altitude of $1425$~m above sea level. The geomagnetic field is defined as $\vec{B}=(F\cos I,\,0,\,-F\sin I)$, where $F=23.424\,\mu\mathrm{T}$ denotes the field strength and $I=-37.42^\circ$ the inclination angle. Longitudinal profiles were recorded at $510$ observation levels spanning $X_v=10-900\,{\rm g\,cm^{-2}}$ in vertical depth.

Radio signals were simulated using the ZHS algorithm implemented in ZHAireS. Time-domain electric fields were recorded with $1\,$ns sampling in a window from $-50\,$ns to $+100\,$ns. 
The longitudinal distributions of electrons and positrons (\textit{Long~$e^{\pm}$}) were exported for comparison with those extracted with radio interferometry. An idealized square antenna grid was simulated, with stations spaced every $100\,$m covering a range of $\pm500\,$m in ground coordinates $(x,y)$. Each antenna time trace was Fourier-transformed to the frequency domain and bandpass filtered to $30-80\,$MHz, matching the frequency range of the AERA and AugerPrime Radio Detector~\cite{PierreAuger:2025kym}. A reference antenna located at $(x,y)=(0,0)$~m was used to define the global timing, and relative offsets were corrected so that all antenna traces share a common time origin prior to interferometric alignment.


\section{Methodology}

\par RIT exploits the radio signals recorded across multiple antennas to reconstruct the coherent pattern of the radio emission from the EAS. By combining the signals, one builds a coherence map that reflects how the radio emission evolves along the shower axis, tracing the longitudinal development of the shower~\cite{RIT1}. In particular, the depth at which the maximum coherence occurs as reconstructed with this method, $X_{\text{RIT}}$, can be compared to the fluorescence-derived measurement of $\Xmax$\,\cite{RIT2}, providing a high-duty-cycle alternative for composition-sensitive observables. It was found in\,\cite{RIT1} that $X_{\text{RIT}}$ does not necessarily coincide with $\Xmax$.

\par The RIT-based reconstruction is performed by scanning a grid of hypothetical emission points $\vec{j}$ arranged in a two-dimensional map, where the vertical axis corresponds to the altitude along the shower axis $\vec{v}$, while the horizontal axis corresponds to the projection of $\vec{v}\times\vec{B}$ onto the ground plane. The accurate reconstruction of the shower direction using RIT has been demonstrated in\,\cite{RIT1,RIT2} and falls outside the scope of this work. Therefore, we will assume to have knowledge of the EAS geometry when constructing the coherence maps. For each individual emission point $\vec{j}$ in the 2-D plane, the expected arrival time of the signal at the antenna $i$ is calculated as
\begin{equation}
\Delta_{i,j} = \frac{d_{i,j}\,\overline{n_{i,j}}}{c},
\end{equation}
where $d_{i,j}$ is the distance between the location of the antenna $i$ and the point $\vec{j}$; $\overline{n_{i,j}}$ is the refractive index averaged along the path of the signal and consistent with the atmospheric model used in ZHAireS; and $c$ is the speed of light.

\par The time traces recorded by the antennas denoted as $S_i(t)$ 
are linearly interpolated to obtain the shifted signals $S_i(t-\Delta_{i,j})$, which are then summed to obtain the expected coherent signal from point $\vec{j}$ denoted as $B_j(t)$ and given by
\begin{equation}
B_j(t)=\sum_{i}^{n_{\rm ant}} S_i(t-\Delta_{i,j}).
\end{equation}

\par
\begin{figure}[!h]
 \centering
 \includegraphics[width=0.9\linewidth]{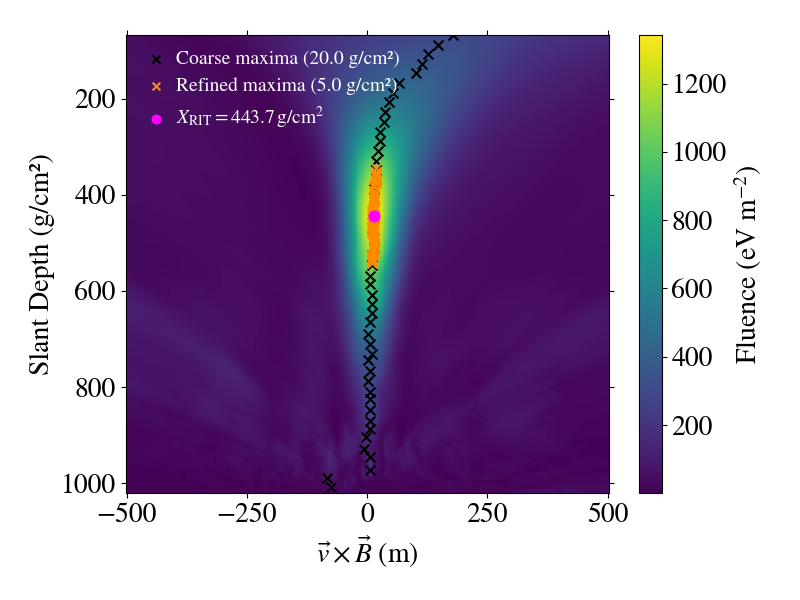}
 \caption{Coherence fluence map in the 2-dimensional plane corresponding to the depth along the shower axis and the direction $\vec{v}\times \vec{B}$ with $\vec{v}$ parallel to the shower axis. The map is obtained for a $1\,$EeV proton-induced shower at $30^\circ$ zenith and $45^\circ$ azimuth simulated with SIBYLL 2.3d hadronic interaction model.}
 \label{fig:1stSkymap}
\end{figure}

 The energy fluence associated with point $\vec{j}$ of the 2D grid is defined as
\begin{equation}
f_{B_j} = \epsilon_0\, c\, \Delta t 
\sum_{t_{\rm peak}-50\text{ns}}^{t_{\rm peak}+50\text{ns}} B_j^2(t),
\end{equation}
where $\Delta t=1\,$ns is the sampling interval and $t_{\rm peak}$ corresponds to time of the maximum of $B^2_j(t)$. 
Repeating this procedure across the grid of points $\vec{j}$ produces a coherence map (an example is shown in Fig.~\ref{fig:1stSkymap}). In this work, the transverse $\vec{v}\times\vec{B}$ direction spans from $-500$~m to $500$~m, while along $\vec{v}$ it extends from an altitude of $1425$~m above sea leval up to $20$~km. Each axis is discretized into $250$ bins, resulting in a high-resolution spatial grid that corresponds to the set of points $\vec{j}$ where the calculations are performed. The fluence itself is computed only once, and the resulting map, initially defined in terms of geometric altitude, is used throughout the analysis.

The longitudinal pattern of the coherent radio emission is obtained by taking the maximum fluence of successive slant-depth slices of the interferometric map. Since the interferometric reconstruction is naturally performed in height coordinates, the height of each layer is converted to slant depth along the shower axis using the Malargüe atmospheric model. Then the accessible slant-depth range of the map is identified, and a set of sampling points is defined with a coarse spacing of $20\,\mathrm{g\,cm^{-2}}$. Since these sampling points do not generally coincide with the discrete heights where the fluence was initially calculated, each depth is assigned to its closest layer. The maximum fluence is then extracted from each layer. To improve the resolution around the region of maximum coherence, a refined sampling of $5\,\mathrm{g\,cm^{-2}}$ is subsequently performed within a window of $\pm 200\,\mathrm{g\,cm^{-2}}$ around the slant depth at which the maximum fluence is obtained, following the same procedure as in the coarser grid. 

For each event, the maximum values of the fluence at each slant-depth slice are used to fit a Gaisser--Hillas (GH) function to determine the depth of maximum coherence $X_{\rm RIT}$. The fit is performed in a region $\pm\,200\,{\rm g\,cm^{-2}}$ around $X_{\rm RIT}$. The fit is expressed in terms of the shifted depth $X' \equiv X - X_{\rm RIT}$ and the normalized fluence, denoted for simplicity as $N' \equiv N/N_{\rm RIT}$. 
In this work, we adopt the normalized Gaisser-Hillas (GH) parameterization in~\cite{Andringa:2011zz} defined as
\begin{equation}
    N' = \left(1+\frac{RX'}{L}\right)^{R^{-2}}
    \exp\left(-\frac{X'}{LR}\right) .
    \label{eq:Gaisser-Hillas}
\end{equation}

The event-by-event Gaisser--Hillas fit reliably extracts first the normalization $N_{\rm RIT}$ and the depth of the maximum $X_{\rm RIT}$, which are used to define the normalized variables $N'$ and $X'$. After this, all profiles are mapped onto a common grid in $X'$ space. Each individual profile is interpolated onto this grid, an example is shown in Fig.\,\ref{fig:ap_USP_RIT&ep_fit} in the Appendix. Finally, the average profile is constructed from the individual profiles and fitted to a GH as explained in the next section. 

\section{Average Universal Shower Profile}
\label{sec:avg_USP_profile}
After extracting the shower maximum and its depth for each event using both RIT and the longitudinal distributions of electrons and positrons, \textit{Long~$e^{\pm}$}, the events within each dataset can be superimposed in the normalized $(X^\prime,N^\prime)$ space. Averaging these normalized profiles yields the corresponding average longitudinal profile.

Prior to constructing the average profile, quality cuts were applied to remove a small fraction of events with unreliable radio-based reconstructions. First, a cut was applied requiring a single peak in the longitudinal profile of the coherent emission. The identification of secondary peaks is based on a peak-finding algorithm, where a candidate peak must satisfy several criteria relative to the global maximum of the profile. In particular, a secondary peak is required to have a height of at least $50\%$ of the maximum value, a minimum prominence of $15\%$ of the maximum, ensuring that it stands out from the surrounding structure, and a separation of at least four sampling bins from any other peak. Most events typically exhibit a secondary, lower-amplitude peak at significantly larger depths than the primary maximum. This feature, while frequent, becomes a problem when the lower-amplitude peak rivals the main peak, and the conditions described above ensure that only these events are being rejected. This effect likely arises from the interferometric reconstruction itself introducing spurious secondary maxima, arising from constructive interference of the electromagnetic signals recorded by the radio antennas. This effect is illustrated in Fig.\,\ref{fig:1stSkymap}, where an enhanced-fluence region appears at large depths and far from the shower axis, where no physical emission is expected. Around $\sim 700\,\mathrm{g\,cm^{-2}}$ and $\sim -300$ m in the $\vec{v}\times\vec{B}$ coordinate, a slight excess above background is visible. This is also apparent in the right tail of the RIT-reconstructed profile in Fig.\,\ref{fig:ap_USP_RIT&ep_fit} in the Appendix. In some events, this effect induces a secondary peak comparable to the main maximum, leading to rejection. Here, however, the peak is too small and narrow to satisfy the secondary peak rejection criteria described in detail above, and the event is, therefore, included in the average profile.

Despite this selection, some events exhibiting strongly non-Gaisser--Hillas  shape remained. These events correspond to showers with very large $\Xmax$, developing deep in the atmosphere and therefore too close to the antenna array for accurate interferometric reconstruction. To remove these cases, an additional cut on $X_{\rm RIT} < $ 750~g\,cm$^{-2}$ was applied effectively rejecting showers with $\Xmax$ close to the ground. Only the events passing these selections were used to construct the average  with a fraction of $\sim 10\%$ rejected. Table~\ref{tab:rit_eff} summarizes the resulting dataset statistics and the impact of the applied cuts.

\begin{table}[h!]
\centering
\begin{tabular}{c|c|c|c|c}
\hline
Dataset&Total&Peak Cut&$X_{\rm RIT}$ Cut &Remaining (\%)\\
\hline
\textit{SIB Proton} & 100 & 9 & 2 & 89.0 \\
\textit{SIB Iron} & 100 & 0 & 0 & 100.0 \\
\textit{QGS Proton} & 100 & 7 & 5 & 88.0 \\
\textit{QGS Iron} & 100 & 0 & 0 & 100.0 \\
\hline
\end{tabular}
\caption{Events simulated for different primaries and hadronic interaction models, together with those surviving the quality cuts, first requiring a single peak in the radio-reconstructed longitudinal profile of coherent emission, and then imposing $X_{\rm RIT}<750$~g\,cm$^{-2}$, as well as the corresponding fraction of events retained. “SIB” denotes the hadronic interaction model SIBYLL~2.3d, and “QGS” refers to QGSJET-II-04.}
\label{tab:rit_eff}
\end{table}

\begin{figure}[!h]
 \centering

\includegraphics[width=0.9\linewidth]{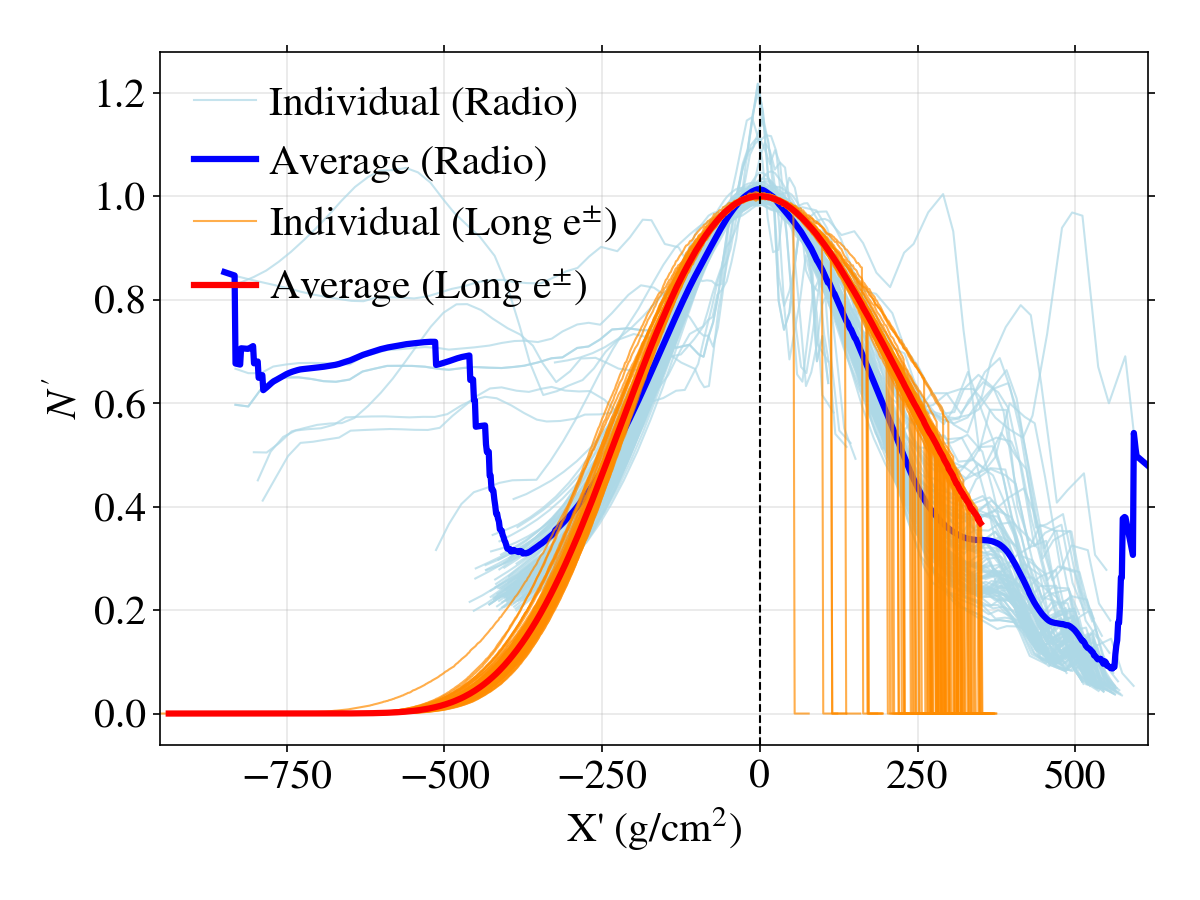}%
\\
\includegraphics[width=0.9\linewidth]{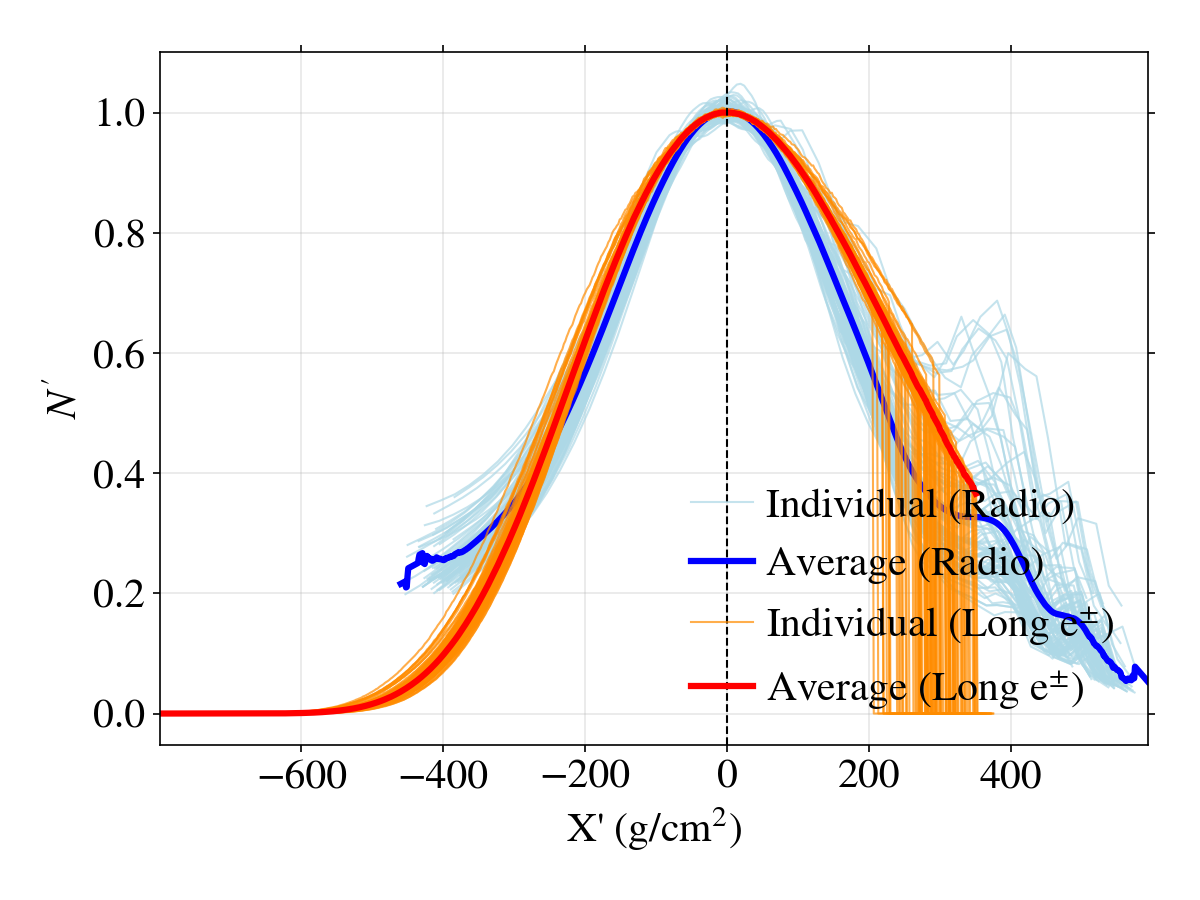}%
 \caption{Average USP for $e^\pm$ and the reconstructed longitudinal patterns of the coherent radio emission. In the top panel, for the individual events (light blue) and the average profile (blue) without applying cuts, in the bottom, after cuts. In orange the individual longitudinal distributions of electrons and positrons directly extracted from ZHAireS, \textit{Long~e$^{\pm}$} and in red their average.}
 \label{fig:average_profs}
\end{figure}

The impact of these cuts is further illustrated in Fig.~\ref{fig:average_profs}. Before applying them, the average radio-derived profile for the \textit{SIB Proton} dataset exhibits a clear deviation from a GH-like shape (top panel). After the cuts are applied, the profile displays a well-defined peak consistent with the expected shower profile (bottom panel).

The radio-derived profiles exhibit larger fluctuations than the longitudinal $e^{\pm}$ distributions, particularly at large depths. In addition, RIT-derived profiles appear systematically shifted higher in the atmosphere with respect to the longitudinal distributions of $e^{\pm}$, a feature of the interferometric reconstruction previously reported in~\cite{RIT1, RIT2}. 

It can also be seen in Fig.\,\ref{fig:average_profs} that, after the cuts, the radio-derived profiles lack data points at small depths, effectively truncating the early-development tail. Conversely, the \textit{Long~$e^{\pm}$} profiles terminate at the observation level for the simulated zenith angle $\theta=30^\circ$, leading to a truncation of the high-depth tail.

\begin{figure}[!h]
 \centering
\includegraphics[width=0.9\linewidth]{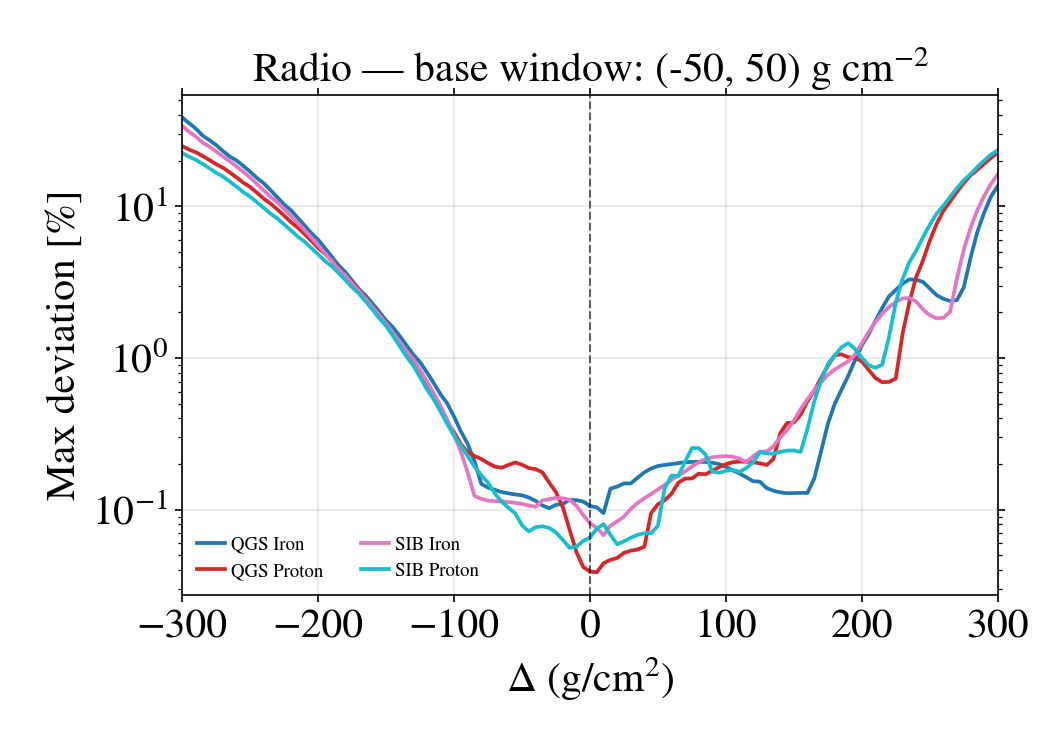}%
\\
\includegraphics[width=0.9\linewidth]{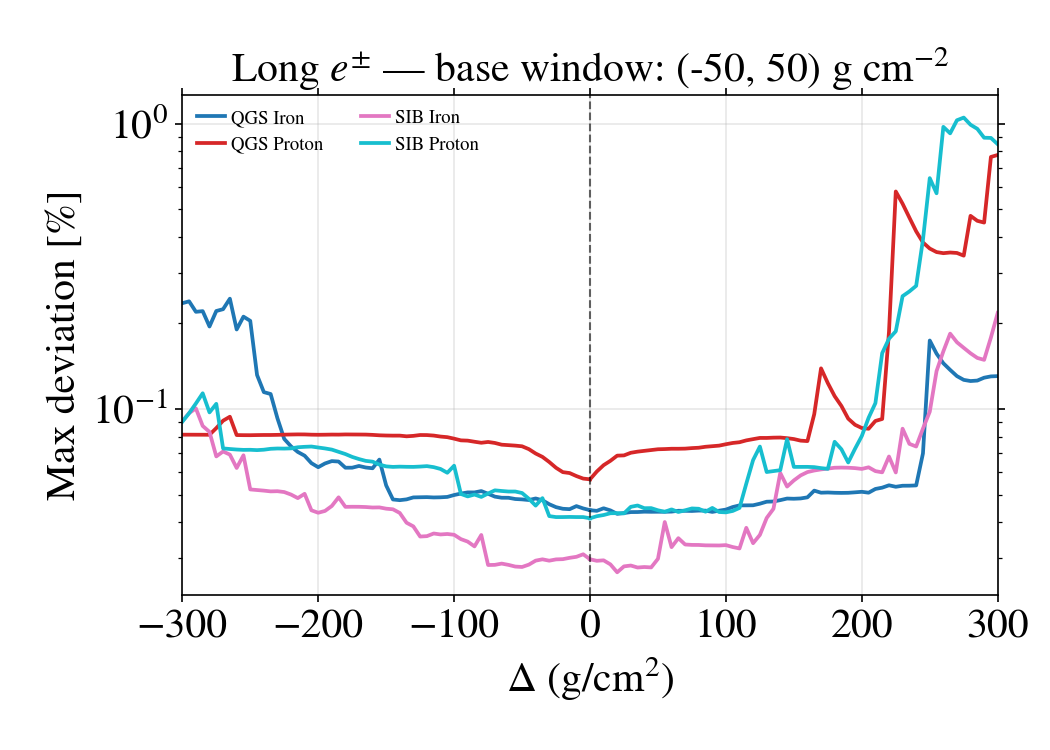}%
 \caption{Maximum difference between the RIT-derived average profile and its corresponding GH fit (top), and between the average \textit{Long~e$^{\pm}$} profile and its corresponding GH fit (bottom), within the fit window defined, in each branch, as [-50, $\Delta$]~g\,cm$^{-2}$ for $\Delta > 0$ and [$\Delta$, 50]~g\,cm$^{-2}$ for $\Delta < 0$, for the analyzed datasets.
}
 \label{fig:window_fits}
\end{figure}

To account for these effects and ensure a reliable Gaisser--Hillas fit, the fitting window was defined requiring that the maximum relative difference between the simulated profile and the GH parameterization remains below $1\%$ within the selected region. A scan of the fitting-window boundaries was performed for all datasets. Starting from a base window of $[-50,\,50]\,{\rm g\,cm^{-2}}$ centered on the maximum, the upper and lower boundaries were extended independently. A parameter $\Delta$ in ${\rm g\,cm^{-2}}$ describes this scan, with $\Delta>0$ corresponding to an extension toward higher depths and $\Delta<0$ toward lower depths. The resulting maximum deviations from the GH are shown in Fig.\,\ref{fig:window_fits} for both the RIT (top) and \textit{Long~$e^{\pm}$} (bottom) average longitudinal profiles.

A stable region consistent with a GH shape is observed around the shower maximum, i.e., near the normalized depth $X'=0$. As expected, apart from the truncation of the high-depth tail caused by the observation level, the \textit{Long~$e^{\pm}$} average profiles (bottom) closely follow the expected USP shape\,\cite{PierreAuger:2018gfc}, with maximum deviations of the order of $0.1\%$, well below the $1\%$ threshold. The RIT-derived average profiles also exhibit a clear USP-like behavior in the vicinity of the shower maximum, though the deviations increase more rapidly outside this region due to the interferometric truncation discussed above.

Based on these results, fixed fitting windows were selected to ensure that the GH parameterization remains within the $1\%$ deviation criterion for all datasets. For the RIT-derived profiles, the maximum deviation exceeds $1\%$ outside the interval $\Delta < -150$~g\,cm$^{-2}$ and $\Delta > 200$~g\,cm$^{-2}$, leading us to adopt a fitting window of [$-200$, $250$]~g\,cm$^{-2}$ ([$-50+\Delta_- $, $50+\Delta_+$]~g\,cm$^{-2}$). A more quantitative description on how the fitting window was chosen can be found in Fig.\,\ref{fig:ap_avg_maxdev_map} in the Appendix. For the \textit{Long~$e^{\pm}$} profiles, a window of [$-300$, $200$]~g\,cm$^{-2}$ was chosen.
To ensure a consistent comparison and a reliable determination of the shape parameters $R$ and $L$, the same fitting procedure is applied to all datasets.

The chosen fitting window excludes the secondary peak commonly observed in RIT profiles, ensuring a robust determination of the shape parameters. The origin of this secondary peak is not yet understood; however, tests performed with alternative fitting ranges show that its presence does not affect the extracted shape parameters within uncertainties. A detailed investigation of this feature is beyond the scope of the present work.

\begin{figure}[!h]
 \centering
\includegraphics[width=0.9\linewidth]{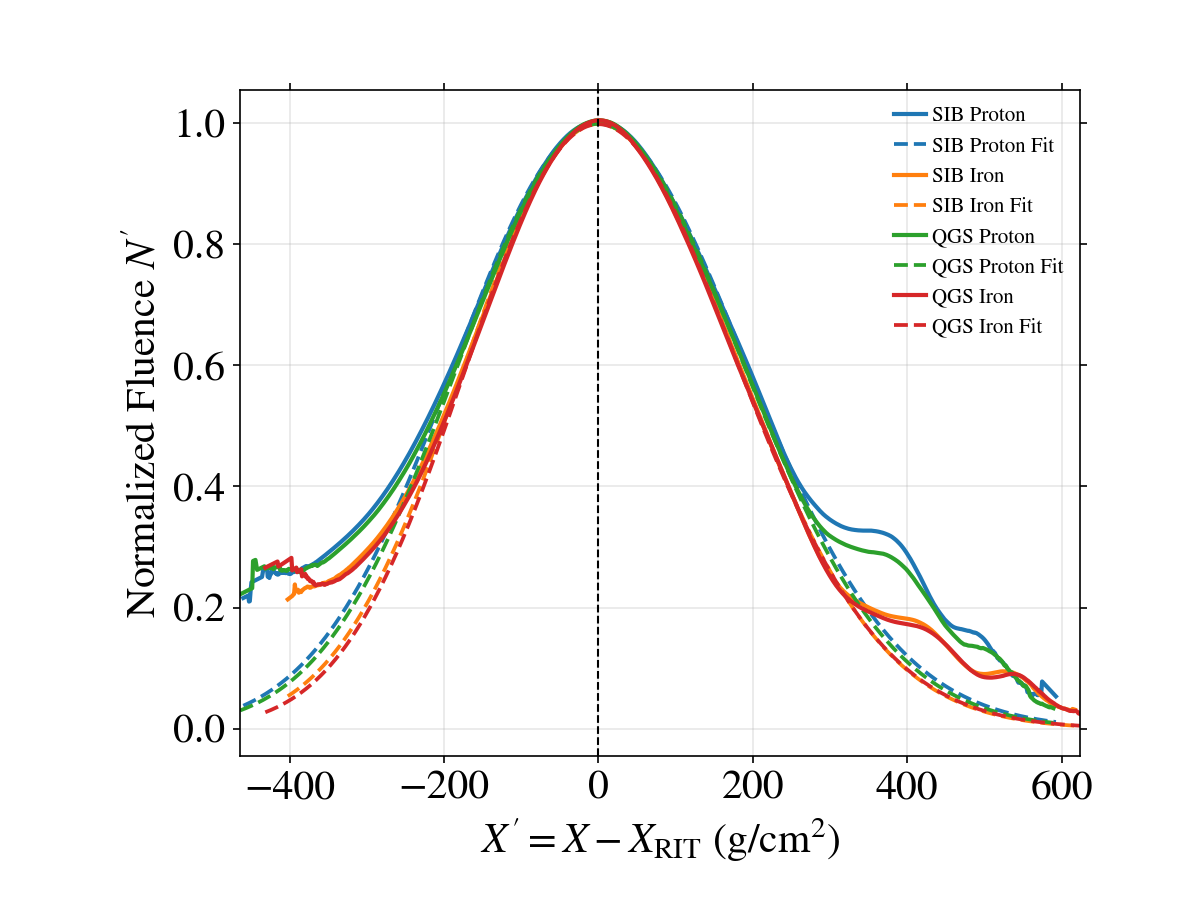}%
\\
\includegraphics[width=0.9\linewidth]{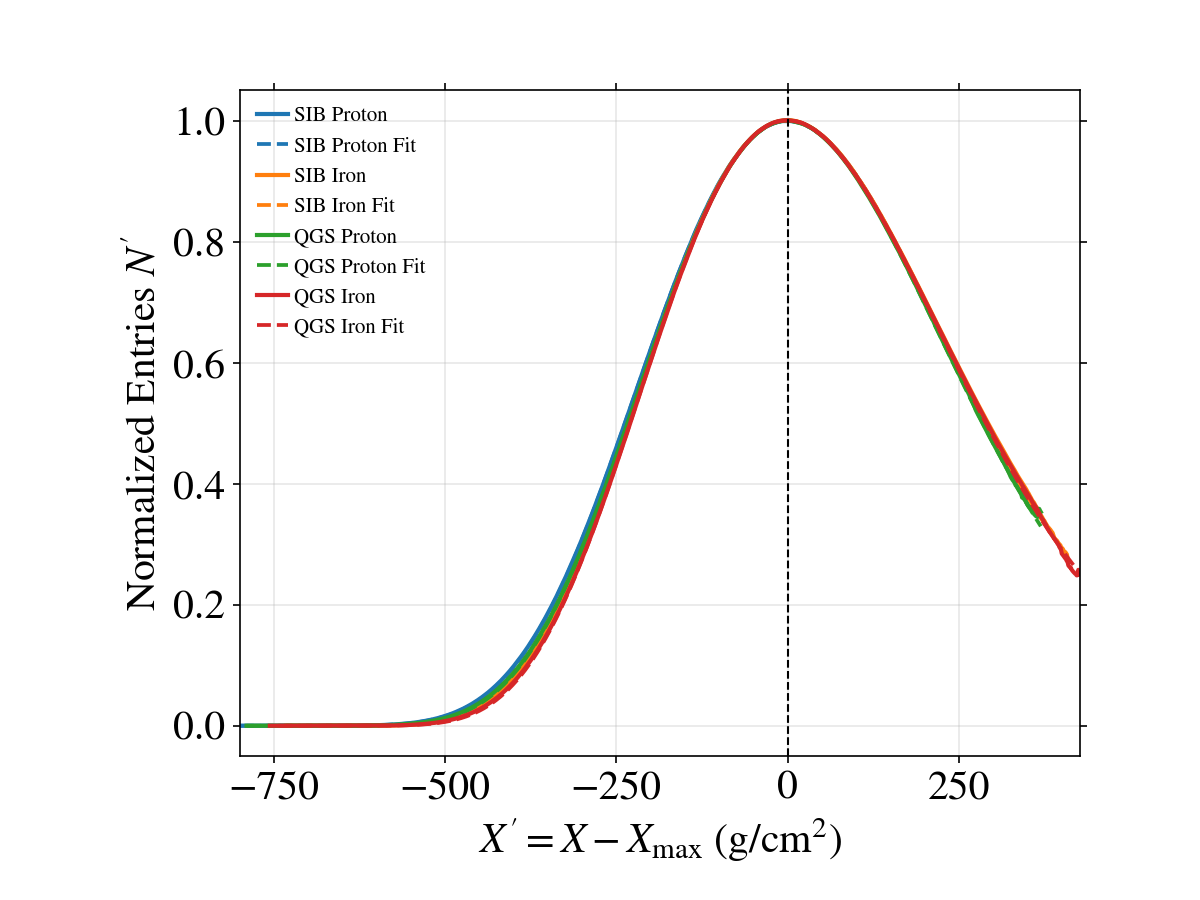}%
 \caption{Normalized average profiles for all datasets, extracted with RIT (top) and \textit{Long~e$^{\pm}$} (bottom), along with their respective GH fits.}
 \label{fig:avg_profiles_and_fits}
\end{figure}

The resulting average profiles and their corresponding GH fits in the normalized $(X^\prime,N^\prime)$ space are shown in Fig.~\ref{fig:avg_profiles_and_fits}. The RIT-derived profiles are presented in the upper panels, while the \textit{Long~$e^{\pm}$} average profiles are shown in the lower panels.
Although RIT-derived average profiles present a GH-like region around their maximum, they are only approximately GH (see Figs.\,\ref{fig:ap_USPvsfitratio} and \ref{fig:ap_USPvsfitratio_th50} in the Appendix for further details), unlike the \textit{Long~$e^{\pm}$} average profiles, which are accurately described by this parametrization. A strong correlation between $\Xrit$ and $\Xmax$ is found as shown in Figs.\,\ref{fig:ap_XRIT_vs_Xmax_alldatasets} and \ref{fig:ap_XRIT_vs_Xmax_diff_zenith} in the Appendix.

With the fitting procedure completed, the average USP shape parameters $R$ and $L$ were extracted for each dataset. First, their dependence on the shower azimuth angle $\phi$ was investigated to ensure that the subsequent analysis in terms of primary mass composition and hadronic interaction models is not biased by the specific geometrical and magnetic field configurations used ($\phi = 45^\circ$ and geomagnetic field inclination $I=-37.42^\circ$).

To this end, additional datasets were generated with ZHAireS using proton primaries and the SIBYLL~2.3d hadronic interaction model. These simulations differ from the \textit{SIB Proton} dataset in the shower azimuth angle and the orientation of the geomagnetic field. Air showers were simulated for discrete azimuth angles between $\phi \in [0^\circ$, $270^\circ]$. In addition, the geomagnetic field inclination was changed ($I=-37.42^\circ$) to $I=-90^\circ$, corresponding to a field perpendicular to the ground. This configuration ensures that the dominant emission component $\vec{v}\times\vec{B}$ and the coherence-map axis are symmetric around $\vec{B}$, so that any differences between datasets should arise primarily from statistical fluctuations.

\begin{figure}[!h]
 \centering
\includegraphics[width=0.95\linewidth]{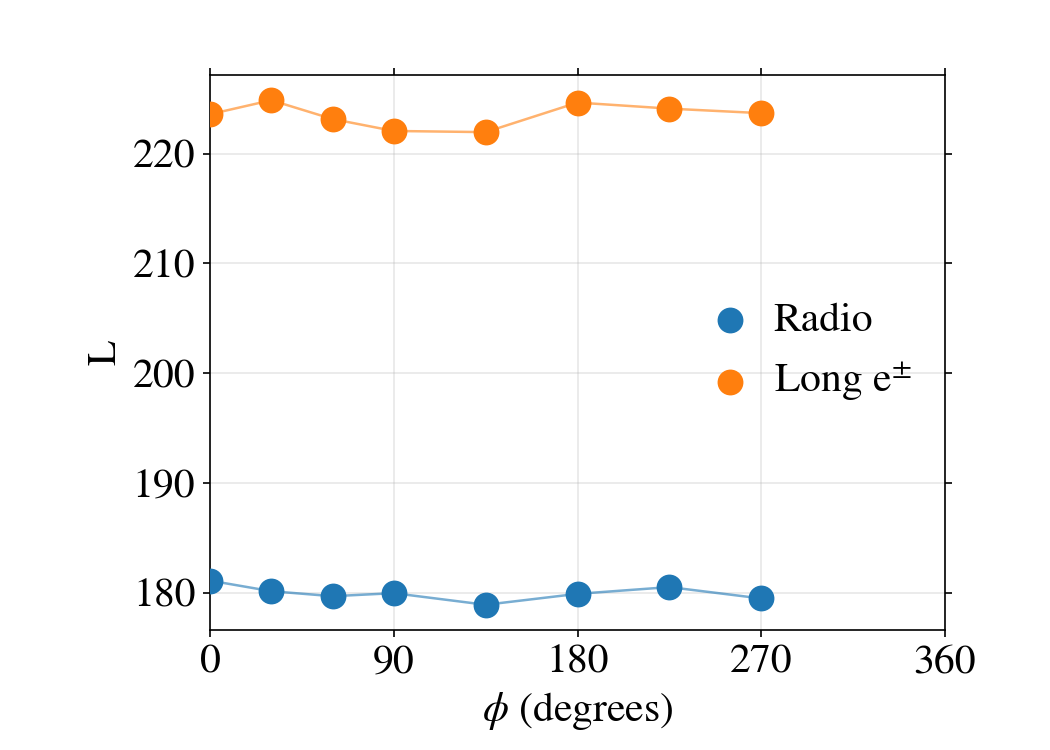}%
\\
\includegraphics[width=0.95\linewidth]{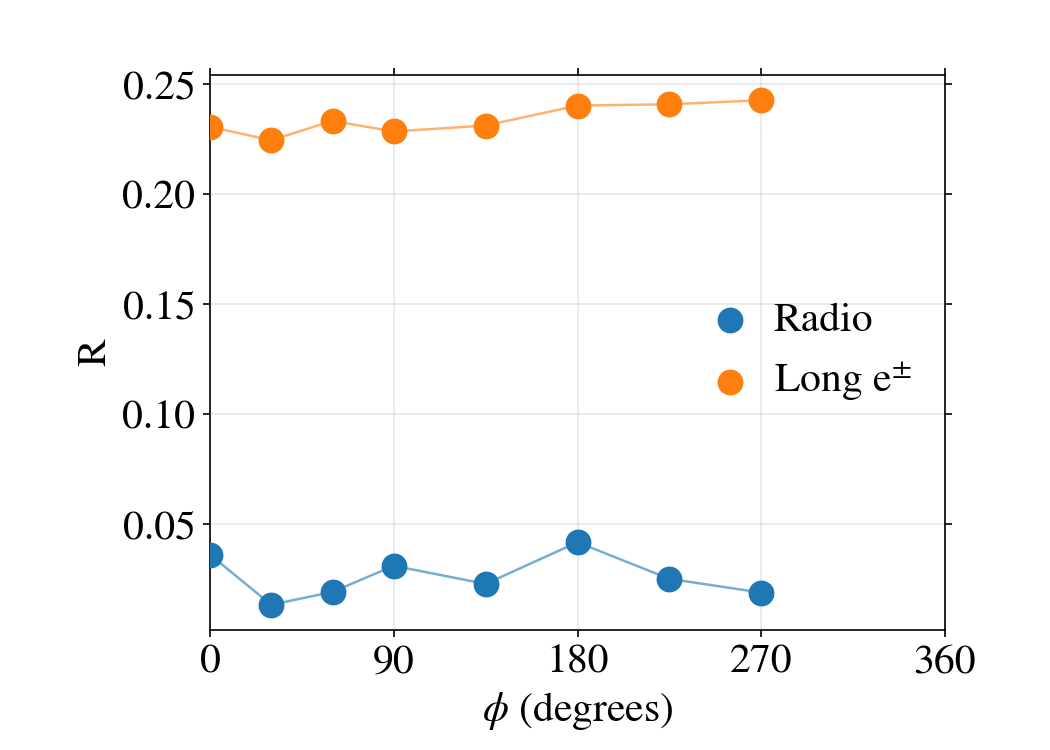}%
 \caption{$L$ (top) and $R$ (bottom) dependence on shower azimuth for the proton simulations performed with the SIBYLL~2.3d hadronic interaction model and a geomagnetic field inclination $I=-90^\circ$.}
 \label{fig:LR_phi}
\end{figure}

Applying the reconstruction and fitting procedure described above to these datasets yields the results shown in Fig.\,\ref{fig:LR_phi}. The $(L,R)$ shape parameters extracted show small variations with the azimuth angle. These fluctuations are slightly larger for the RIT-derived profiles than for the \textit{Long~$e^{\pm}$} profiles, particularly for the parameter $R$, but remain consistent with statistical uncertainties.

\section{Average USP shape dependence on hadronic models and primary mass}

The sensitivity of the RIT-derived average USP to the primary mass composition and the hadronic interaction model is assessed through the Gaisser–Hillas shape parameters, $L$ and $R$, extracted previously. The results obtained from the radio-derived profiles are compared with those obtained directly from the longitudinal distributions of electrons and positrons (\textit{Long~$e^{\pm}$}), which provide a benchmark equivalent to those reconstructed with fluorescence detectors.

The statistical stability of the extracted shape parameters was assessed using a \textit{leave-one-out} resampling (\textit{jackknife}) procedure. For each event in a given dataset, the average profile was reconstructed from the remaining events and the GH fitting and parameter extraction were repeated. This procedure generates a distribution of points in the $(R,L)$ parameter space around the values obtained from the average profile constructed using the full dataset.

\begin{figure}[!h]
 \centering
\includegraphics[width=0.9\linewidth]{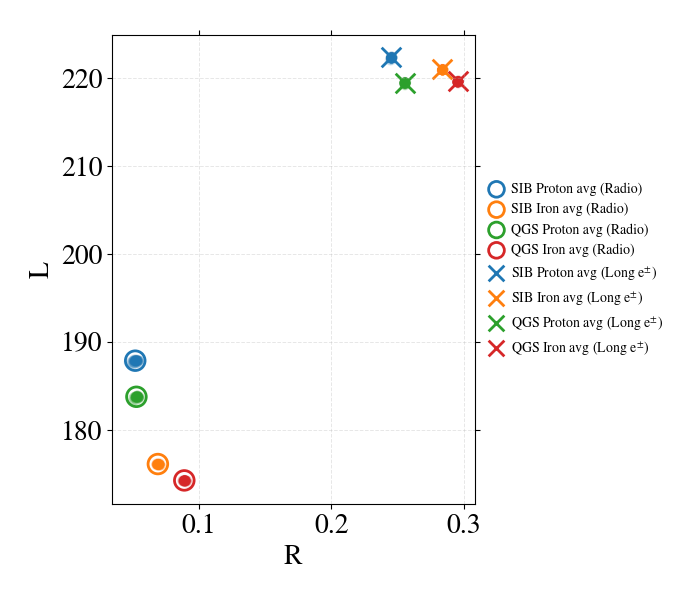}%
 \caption{Shape parameters $(R,L)$ derived from GH fits to average normalized profiles. Open circles: radio-derived; crosses: \textit{Long~$e^{\pm}$}. Proton (blue/green) and iron (orange/red) datasets are clearly separated. The cloud of points within the open circles and overlapping the crosses represents the distribution of shape parameters obtained from the same dataset (indicated by color), where each point is computed by removing a different event from the dataset before averaging.}
 \label{fig:Umbrella_LR}
\end{figure}

The results are shown in Fig.~\ref{fig:Umbrella_LR}. The shape parameters $R$ and $L$ obtained from the full datasets are indicated by empty circles for the RIT reconstruction and by crosses for the \textit{Long~$e^{\pm}$} benchmark. Filled circles show the jackknife resampling results, forming a cloud of points around the full-dataset estimates that reflects the statistical stability of the extracted parameters\footnote{An additional dataset of $1000$ proton showers, generated with SIBYLL~2.3d (ten times larger than \textit{SIB Proton}), was used to assess statistical stability. Consistent with the \textit{jackknife} results, the deviation in the $(R,L)$ space between datasets with different statistics is $0.60\%$, well below the $2.18\%$ separation between \textit{SIB Proton} and \textit{QGS Proton}.}. 

Fig.\,\ref{fig:Umbrella_LR} shows that the RIT-derived average USP provides a clear separation between proton and iron primaries, with a discrimination power comparable to that obtained from fluorescence-based profiles. A measurable sensitivity to the hadronic interaction model is also observed. When comparing the RIT-derived parameters with those obtained with the \textit{Long~$e^{\pm}$} distributions, the datasets appear shifted, clustering around distinct regions of the $(R,L)$ parameter space. 

A notable feature of the RIT reconstruction is an enhanced sensitivity to $L$, while the sensitivity to $R$ remains comparable to that obtained from the fluorescence-equivalent benchmark.

\begin{figure}[!h]
 \centering

\includegraphics[width=0.95\linewidth]{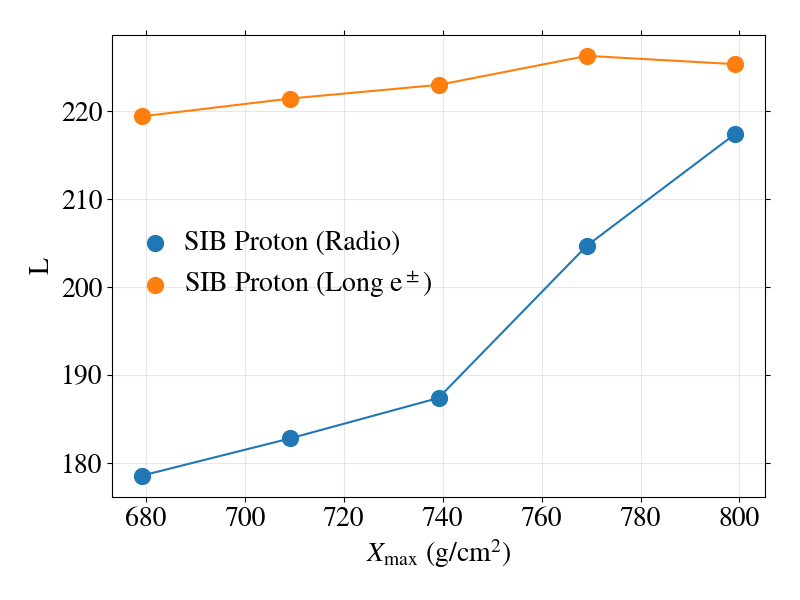}%
\\
\includegraphics[width=0.95\linewidth]{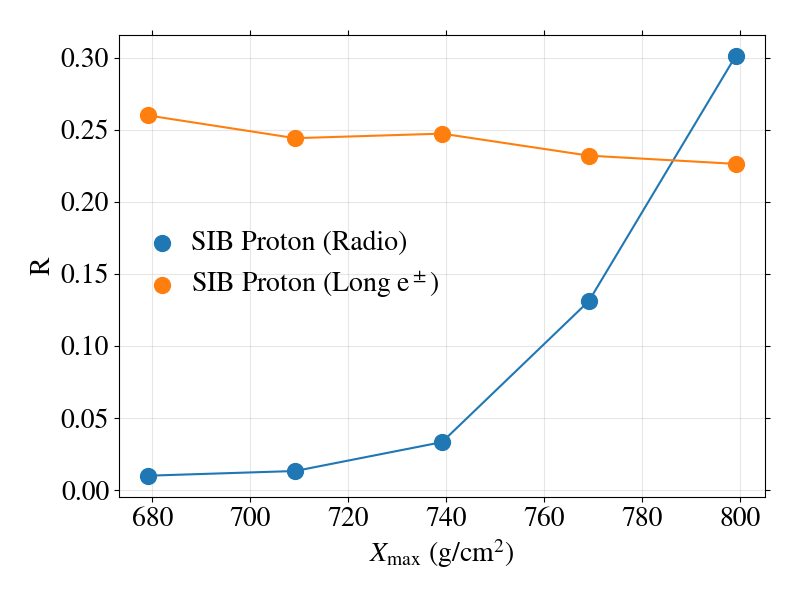}%
 \caption{L (top) and R (bottom) dependence on $X_{\rm max}$ for the \textit{SIB Proton} dataset.}
 \label{fig:LR_vs_Xmax}
\end{figure}

To investigate the origin of this behavior, the datasets were divided into bins of $\Xmax$, allowing the dependence of the shape parameters on the depth of the shower maximum to be studied. For each bin, an average profile was constructed and fitted with the GH parameterization. The resulting values of $R$ and $L$ are shown in Fig.\,\ref{fig:LR_vs_Xmax}. A clear dependence on $\Xmax$ is observed for the RIT-derived parameters, particularly for $L$ (top panel), and to a lesser extent for $R$ (bottom panel). In contrast, the \textit{Long~$e^{\pm}$} profiles show little variation with $\Xmax$. The origin of this behavior in the RIT-derived profiles is not yet fully understood and will require further investigation in future work.

However, the observed dependence provides an explanation for the enhanced composition sensitivity seen in Fig.\,\ref{fig:Umbrella_LR}. Since iron-induced showers reach their maximum at smaller $\Xmax$ than proton-induced showers, the dependence of the parameter $L$ on $\Xmax$ amplifies the separation between primaries along the $L$ direction in the $(R,L)$ space. 

These results demonstrate that radio interferometry can access the shape parameters of the average USP which can be used to probe both primary mass composition and hadronic interaction models, providing a high-duty-cycle complement to fluorescence measurements.

\section{Summary and Discussion}
\label{sec:conclusions}

In this work we have investigated the capability of radio interferometry to reconstruct the average universal shower profile (USP) of extensive air showers and to extract its shape parameters. Using simulated datasets and an idealized antenna array, we have shown that the average USP can be reconstructed from radio measurements and that the corresponding Gaisser--Hillas shape parameters $(R,L)$ can be reliably extracted.

The extracted shape parameters obtained from the radio-derived profiles exhibit a separation between proton and iron primaries comparable to that obtained from the longitudinal distributions of electrons and positrons, used here as a fluorescence-equivalent benchmark. In addition, a measurable dependence on the hadronic interaction model is observed. These results demonstrate that radio interferometry can access higher-order information on the longitudinal development of air showers beyond $\Xmax$, by exploiting the universal features of the electromagnetic profile encoded in the average USP.

Given the near 100\% duty cycle of radio detectors, in contrast to the much lower duty cycle of fluorescence measurements, the radio interferometric technique provides a complementary approach for large-statistics studies of primary mass composition and hadronic interactions at ultra-high energies.

Future work should focus on extending this study to more realistic detector configurations, including existing radio arrays such as the Auger Radio Detector, and on evaluating the performance of the method for more inclined showers and under realistic detector and noise conditions. These studies will be essential to assess the applicability of the method to experimental data.

\section*{Acknowledgments}

We would like to thank Vladim\'ir Novotn\'y for carefully reading the manuscript and for useful comments.
This work has been partially funded by Fundação para a Ciência e Tecnologia, Portugal, under project \url{https://doi.org/10.54499/2024.06879.CERN}; by
Ministerio de Ciencia, Innovaci\'on y Universidades/Agencia Estatal de Investigaci\'on MICIU/AEI /10.13039/501100011033, Spain
(PID2022-140510NB-I00, PCI2023-145952-2, CNS2024-154676, and Mar\'\i a de Maeztu grant CEX2023-001318-M);
Xunta de Galicia, Spain (CIGUS Network of Research Centers and 
Consolidaci\'on 2021 GRC GI-2033 ED431C-2021/22 and 2022 ED431F-2022/15);
Feder Funds;
and European Union ERDF.

\bibliography{references}

@article{Albrecht:2025kbb,
    author = "Albrecht, J. and others",
    title = "{Global tuning of hadronic interaction models with accelerator-based and astroparticle data}",
    eprint = "2508.21796",
    archivePrefix = "arXiv",
    primaryClass = "astro-ph.HE",
    reportNumber = "FERMILAB-PUB-26-0089-PPD",
    doi = "10.1038/s42254-025-00897-3",
    journal = "Nature Rev. Phys.",
    volume = "8",
    number = "2",
    pages = "98--114",
    year = "2026"
}

@article{PierreAuger:2023lkx,
    author = "Abdul Halim, A. and others",
    collaboration = "Pierre Auger",
    title = "{Demonstrating Agreement between Radio and Fluorescence Measurements of the Depth of Maximum of Extensive Air Showers at the Pierre Auger Observatory}",
    eprint = "2310.19963",
    archivePrefix = "arXiv",
    primaryClass = "astro-ph.HE",
    reportNumber = "FERMILAB-PUB-24-0138-AD-CSAID-PPD-TD-V",
    doi = "10.1103/PhysRevLett.132.021001",
    journal = "Phys. Rev. Lett.",
    volume = "132",
    number = "2",
    pages = "021001",
    year = "2024"
}

@article{DeHenau:2025zoh,
    author = "De Henau, Vital and Bouma, S. and Bray, J. and Buitink, S. and Corstanje, A. and Desmet, M. and Dickinson, E. and van Dongen, L. and Hare, B. and Huege, T.",
    title = "{Investigating double bump air showers with the SKA}",
    eprint = "2510.13788",
    archivePrefix = "arXiv",
    primaryClass = "astro-ph.HE",
    doi = "10.22323/1.501.0236",
    journal = "PoS",
    volume = "ICRC2025",
    pages = 236,
    year = 2025
}

@article{deErrico:2026usf,
    author = "de Errico, Beatriz and Timmermans, Charles",
    title = "{Air shower development through the time dependence of its induced electric field}",
    eprint = {2603.05424},
    archivePrefix = {arXiv},
    primaryClass = {hep-ex},
    journal = {},
    month = 3,
    year = 2026
}

@article{RIT1,
   title={Radio interferometry applied to the observation of cosmic-ray induced extensive air showers},
   volume={81},
   ISSN={1434-6052},
   url={http://dx.doi.org/10.1140/epjc/s10052-021-09925-9},
   DOI={10.1140/epjc/s10052-021-09925-9},
   number={12},
   journal={The European Physical Journal C},
   publisher={Springer Science and Business Media LLC},
   author={Schoorlemmer, Harm and Carvalho, Washington R.},
   year={2021},
   month=dec }

@article{RIT2,
doi = {10.1088/1748-0221/16/07/P07048},
url = {https://doi.org/10.1088/1748-0221/16/07/P07048},
year = {2021},
month = {jul},
publisher = {IOP Publishing},
volume = {16},
number = {07},
pages = {P07048},
author = {Schlüter, F. and Huege, T.},
title = {Expected performance of air-shower measurements with the radio-interferometric technique},
journal = {Journal of Instrumentation}
}

@article{Engel:2011zzb,
    author = "Engel, Ralph and Heck, Dieter and Pierog, Tanguy",
    title = "{Extensive air showers and hadronic interactions at high energy}",
    doi = "10.1146/annurev.nucl.012809.104544",
    journal = "Ann. Rev. Nucl. Part. Sci.",
    volume = "61",
    pages = "467--489",
    year = "2011"
}

@article{PierreAuger:2014gko,
    author = "Aab, A. and others",
    collaboration = "Pierre Auger",
    title = "{Depth of maximum of air-shower profiles at the Pierre Auger Observatory. II. Composition implications}",
    eprint = "1409.5083",
    archivePrefix = "arXiv",
    primaryClass = "astro-ph.HE",
    reportNumber = "FERMILAB-PUB-14-347-AD-AE-CD-TD",
    doi = "10.1103/PhysRevD.90.122006",
    journal = "Phys. Rev. D",
    volume = "90",
    number = "12",
    pages = "122006",
    year = "2014"
}

@article{Kampert:2012mx,
    author = "Kampert, Karl-Heinz and Unger, Michael",
    title = "{Measurements of the Cosmic Ray Composition with Air Shower Experiments}",
    eprint = "1201.0018",
    archivePrefix = "arXiv",
    primaryClass = "astro-ph.HE",
    doi = "10.1016/j.astropartphys.2012.02.004",
    journal = "Astropart. Phys.",
    volume = "35",
    pages = "660--678",
    year = "2012"
}

@article{PierreAuger:2024neu,
    author = "Abdul Halim, A. and others",
    collaboration = "Pierre Auger",
    title = "{Testing hadronic-model predictions of depth of maximum of air-shower profiles and ground-particle signals using hybrid data of the Pierre Auger Observatory}",
    eprint = "2401.10740",
    archivePrefix = "arXiv",
    primaryClass = "astro-ph.HE",
    reportNumber = "FERMILAB-PUB-24-0065-PPD-TD",
    doi = "10.1103/PhysRevD.109.102001",
    journal = "Phys. Rev. D",
    volume = "109",
    number = "10",
    pages = "102001",
    year = "2024"
}

@article{PierreAuger:2021qsd,
    author = "Aab, Alexander and others",
    collaboration = "Pierre Auger",
    title = "{Measurement of the Fluctuations in the Number of Muons in Extensive Air Showers with the Pierre Auger Observatory}",
    eprint = "2102.07797",
    archivePrefix = "arXiv",
    primaryClass = "hep-ex",
    reportNumber = "FERMILAB-PUB-21-202-AD-AE-SCD-TD",
    doi = "10.1103/PhysRevLett.126.152002",
    journal = "Phys. Rev. Lett.",
    volume = "126",
    number = "15",
    pages = "152002",
    year = "2021"
}

@article{PierreAuger:2018gfc,
    author = "Aab, Alexander and others",
    collaboration = "Pierre Auger",
    title = "{Measurement of the average shape of longitudinal profiles of cosmic-ray air showers at the Pierre Auger Observatory}",
    eprint = "1811.04660",
    archivePrefix = "arXiv",
    primaryClass = "astro-ph.HE",
    reportNumber = "FERMILAB-PUB-18-633-ND-TD",
    doi = "10.1088/1475-7516/2019/03/018",
    journal = "JCAP",
    volume = "03",
    pages = "018",
    year = "2019"
}

@article{Conceicao:2015toa,
    author = "Concei{\c{c}}{\~a}o, R. and Andringa, S. and Diogo, F. and Pimenta, M.",
    title = "{The average longitudinal air shower profile: exploring the shape information}",
    doi = "10.1088/1742-6596/632/1/012087",
    journal = "J. Phys. Conf. Ser.",
    volume = "632",
    number = "1",
    pages = "012087",
    year = "2015"
}

@article{Andringa:2011ik,
    author = "Andringa, S. and Cazon, L. and Concei{\c{c}}{\~a}o, R. and Pimenta, M.",
    title = "{The Muonic longitudinal shower profiles at production}",
    eprint = "1111.1424",
    archivePrefix = "arXiv",
    primaryClass = "hep-ph",
    doi = "10.1016/j.astropartphys.2012.03.010",
    journal = "Astropart. Phys.",
    volume = "35",
    pages = "821--827",
    year = "2012"
}

@article{Andringa:2011zz,
    author = "Andringa, S. and Concei{\c{c}}{\~a}o, R. and Pimenta, M.",
    title = "{Mass composition and cross-section from the shape of cosmic ray shower longitudinal profiles}",
    doi = "10.1016/j.astropartphys.2010.10.002",
    journal = "Astropart. Phys.",
    volume = "34",
    pages = "360--367",
    year = "2011"
}

@article{qgsjet,
  author        = {Ostapchenko, Sergey},
  journal       = {Phys. Rev. D},
  title         = {{Monte Carlo treatment of hadronic interactions in enhanced Pomeron scheme: I. QGSJET-II model}},
  year          = {2011},
  pages         = {014018},
  volume        = {83},
  archiveprefix = {arXiv},
  doi           = {10.1103/PhysRevD.83.014018},
  eprint        = {1010.1869},
  primaryclass  = {hep-ph},
  slaccitation  = {%%CITATION = ARXIV:1010.1869;%%},
}

@Article{sibyll,
  author    = {Riehn, Felix and Engel, Ralph and Fedynitch, Anatoli and Gaisser, Thomas K. and Stanev, Todor},
  journal   = {Phys. Rev. D},
  title     = {{Hadronic interaction model SIBYLL 2.3d and extensive air showers}},
  year      = {2020},
  month     = {Sep},
  pages     = {063002},
  volume    = {102},
  doi       = {10.1103/PhysRevD.102.063002},
  issue     = {6},
  numpages  = {28},
  publisher = {American Physical Society},
}

@article{Kahn:1966,
  author = {Kahn, F. D. and Lerche, I.},
  title = {Radiation from cosmic ray air showers},
  journal = {Proc. Roy. Soc. A},
  volume = {289},
  pages = {206},
  year = {1966}
}

@article{Askaryan:1962,
  author = {Askaryan, G. A.},
  title = {Excess negative charge of an electron-photon shower and its coherent radio emission},
  journal = {Sov. Phys. JETP},
  volume = {14},
  pages = {441},
  year = {1962}
}

@article{Huege:2016,
  author = {Huege, Tim},
  title = {Radio detection of cosmic ray air showers in the digital era},
  journal = {Physics Reports},
  volume = {620},
  pages = {1-52},
  year = {2016}
}

@article{Buitink:2016nkf,
    author = "Buitink, S. and others",
    title = "{A large light-mass component of cosmic rays at 10{\textasciicircum}{17} - 10{\textasciicircum}{17.5} eV from radio observations}",
    eprint = "1603.01594",
    archivePrefix = "arXiv",
    primaryClass = "astro-ph.HE",
    doi = "10.1038/nature16976",
    journal = "Nature",
    volume = "531",
    pages = "70",
    year = "2016"
}

@article{Corstanje:2025wbc,
    author = "Corstanje, A. and others",
    title = "{LOFAR-style reconstruction of cosmic-ray air showers with SKA-Low}",
    eprint = "2504.16873",
    archivePrefix = "arXiv",
    primaryClass = "astro-ph.HE",
    doi = "10.1103/l8mt-994v",
    journal = "Phys. Rev. D",
    volume = "112",
    number = "2",
    pages = "023017",
    year = "2025"
}

@article{Bezyazeekov:2018yjw,
    author = "Bezyazeekov, P. A. and others",
    title = "{Reconstruction of cosmic ray air showers with Tunka-Rex data using template fitting of radio pulses}",
    eprint = "1803.06862",
    archivePrefix = "arXiv",
    primaryClass = "astro-ph.IM",
    doi = "10.1103/PhysRevD.97.122004",
    journal = "Phys. Rev. D",
    volume = "97",
    number = "12",
    pages = "122004",
    year = "2018"
}

@article{aires,
  author    = {Sciutto, S. J.},
  title     = {AIRES: A system for air shower simulations},
  journal   = {arXiv preprint},
  eprint    = {astro-ph/9911331},
  year      = {1999}
}

@article{ZHS92,
  author  = {Zas, E. and Halzen, F. and Stanev, T.},
  title   = {Electromagnetic pulses from high-energy showers: Implications for neutrino detection},
  journal = {Phys. Rev. D},
  volume  = {45},
  pages   = {362--376},
  year    = {1992},
  doi     = {10.1103/PhysRevD.45.362}
}

@article{TimeDomainZHS,
  author  = {Alvarez-Muñiz, J. and Romero-Wolf, A. and Zas, E.},
  title   = {Practical and accurate calculations of Askaryan radiation},
  journal = {Phys. Rev. D},
  volume  = {81},
  pages   = {123009},
  year    = {2010},
  doi     = {10.1103/PhysRevD.81.123009}
}

@article{ZHAireS,
  author  = {Alvarez-Muñiz, J. and Carvalho, W. R. and Zas, E.},
  title   = {Monte Carlo simulations of radio pulses in atmospheric showers using ZHAireS},
  journal = {Astroparticle Physics},
  volume  = {35},
  pages   = {325--341},
  year    = {2012},
  doi     = {10.1016/j.astropartphys.2011.10.005}
}

@article{PierreAuger:2025kym,
    author = "Abdul Halim, A. and others",
    collaboration = "Pierre Auger",
    title = "{Measuring the muon content of inclined air showers using AERA and the water-Cherenkov detectors of the Pierre Auger Observatory}",
    eprint = "2507.02558",
    archivePrefix = "arXiv",
    primaryClass = "astro-ph.HE",
    reportNumber = "FERMILAB-PUB-25-0440-PPD-TD",
    doi = "10.1103/2q9f-pbrp",
    journal = "Phys. Rev. D",
    volume = "112",
    number = "12",
    pages = "123042",
    year = "2025"
}

\clearpage
\section*{Appendix}
\appendix

In Fig.~\ref{fig:ap_USP_RIT&ep_fit}, we show an example of a RIT-derived longitudinal profile of the coherent emission (blue dots) for the same event shown in Fig.\,\ref{fig:1stSkymap}, in the normalized $N'$ and $X'$ space, along with the longitudinal profile of e$^\pm$ (orange solid lines), and their GH fits (dashed lines). Note that for the RIT-derived profile, $N'\equiv N/N_{\rm RIT}$, where $N$ represents fluence and $X'\equiv X/X_{\rm RIT}$, while for \textit{Long e$^\pm$} $N'\equiv N/N_{\rm max}$ and $X'\equiv X/X_{\rm max}$. This  event serves as an example of the RIT-originated deformations consistently observed across datasets, since despite its \textit{Long e$^\pm$} profile presenting no significant deviations from a GH function, its RIT-derived longitudinal profile deviates from the expected GH shape away from its maximum. 

\begin{figure}[!h]
 \centering
 \includegraphics[width=0.9\linewidth]{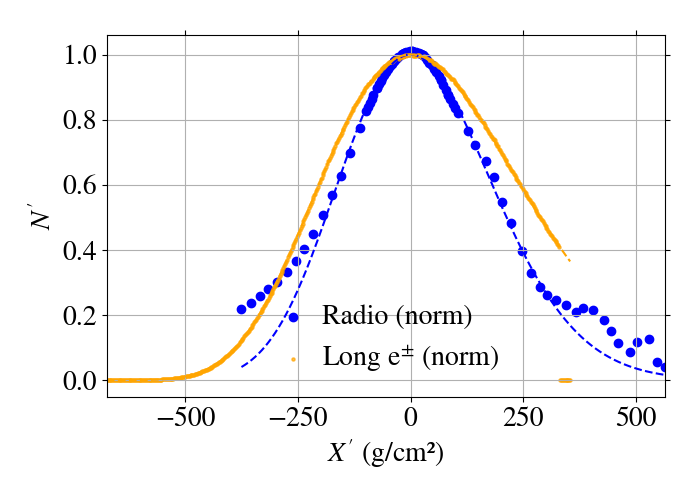}%
 \caption{RIT-derived (blue dots) and \textit{Long e$^\pm$} (orange solid line) longitudinal profiles in the normalized $X'$ and $N'$ space for the same event shown in Fig.~\ref{fig:1stSkymap}. In dashed lines of the same colors, the corresponding GH fits are also shown.}
 \label{fig:ap_USP_RIT&ep_fit}
\end{figure}

In Fig.\,\ref{fig:ap_USPvsfitratio}, we show the RIT-derived average profile for the \textit{SIB Proton} dataset together with its GH fit (see also Fig.\,\ref{fig:avg_profiles_and_fits}, top, blue lines). The ratio plot below highlights the deviation between the fit and the RIT profile, demonstrating good agreement within the window used for extracting the $L$ and $R$ parameters, namely $X' \in [-200, 250]$~g\,cm$^{-2}$. This consistency around the peak supports the approximately GH-like shape of the RIT-derived average longitudinal profile, in line with Fig.~\ref{fig:window_fits}. Under these conditions, the shape parameters $R$ and $L$ can be reliably extracted and compared across datasets.

\begin{figure}[!h]
 \centering
 \includegraphics[width=0.9\linewidth]{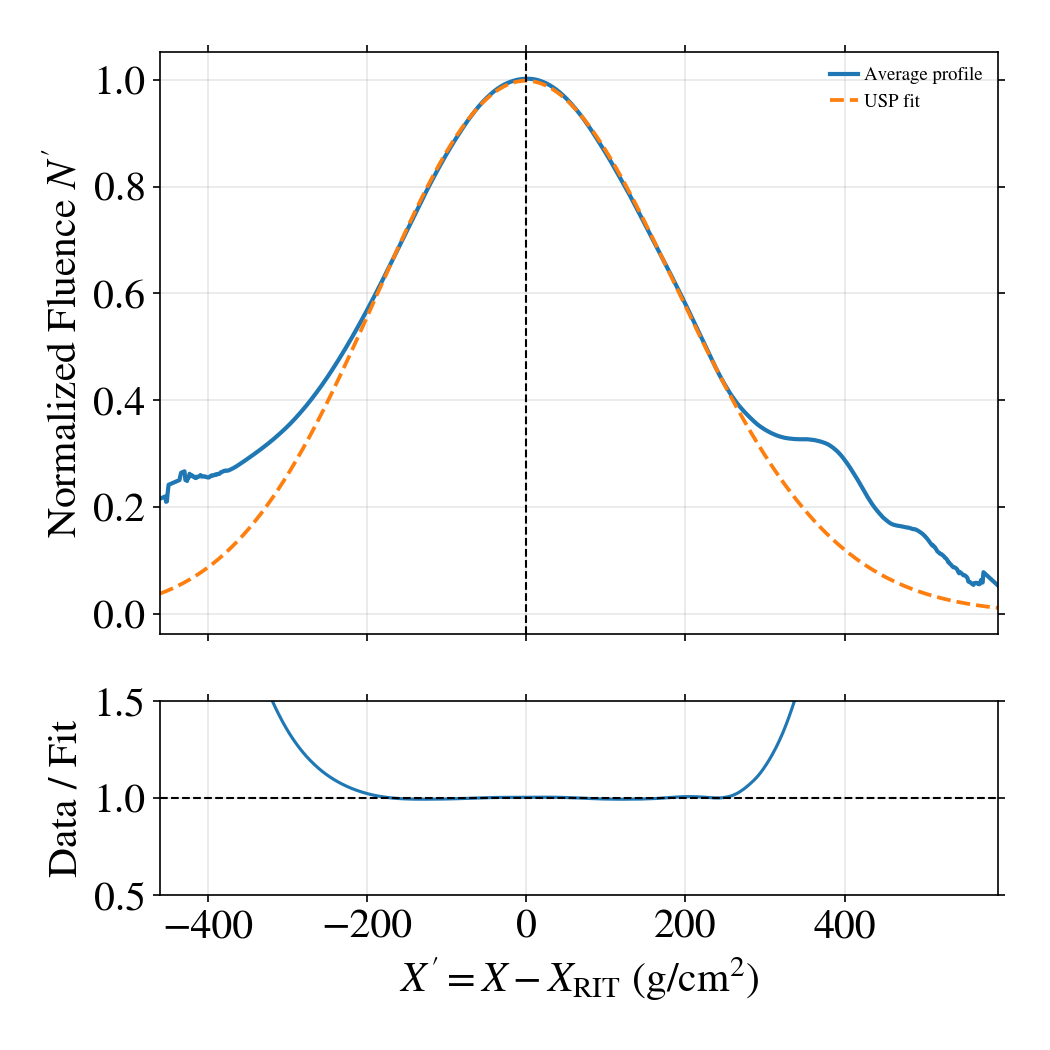}%
 \caption{RIT-derived average profile, applied fit to Eq.\,(\ref{eq:Gaisser-Hillas}), and ratio between them for the \textit{SIB Proton} dataset and zenith angle $\theta=30^\circ$.}
 \label{fig:ap_USPvsfitratio}
\end{figure}

In Fig.\,\ref{fig:ap_USPvsfitratio_th50}, we show the corresponding average RIT-derived profile, GH fit and daviation for the \textit{SIB Proton} $\theta$ = $50^\circ$ dataset, differing from \textit{SIB Proton} only in zenith angle ($\theta=30^\circ$ for \textit{SIB Proton}). At this zenith angle, the discrepancies between the average profile and the fit deviate from those shown in Fig.\,\ref{fig:avg_profiles_and_fits} (top). Nevertheless, the RIT-derived average profile still retains an approximate GH-like shape near $\Xrit$, indicating the validity of the method for higher inclination showers (up to the tested $\theta = 50^{\circ}$).

\begin{figure}[!h]
 \centering
 \includegraphics[width=0.9\linewidth]{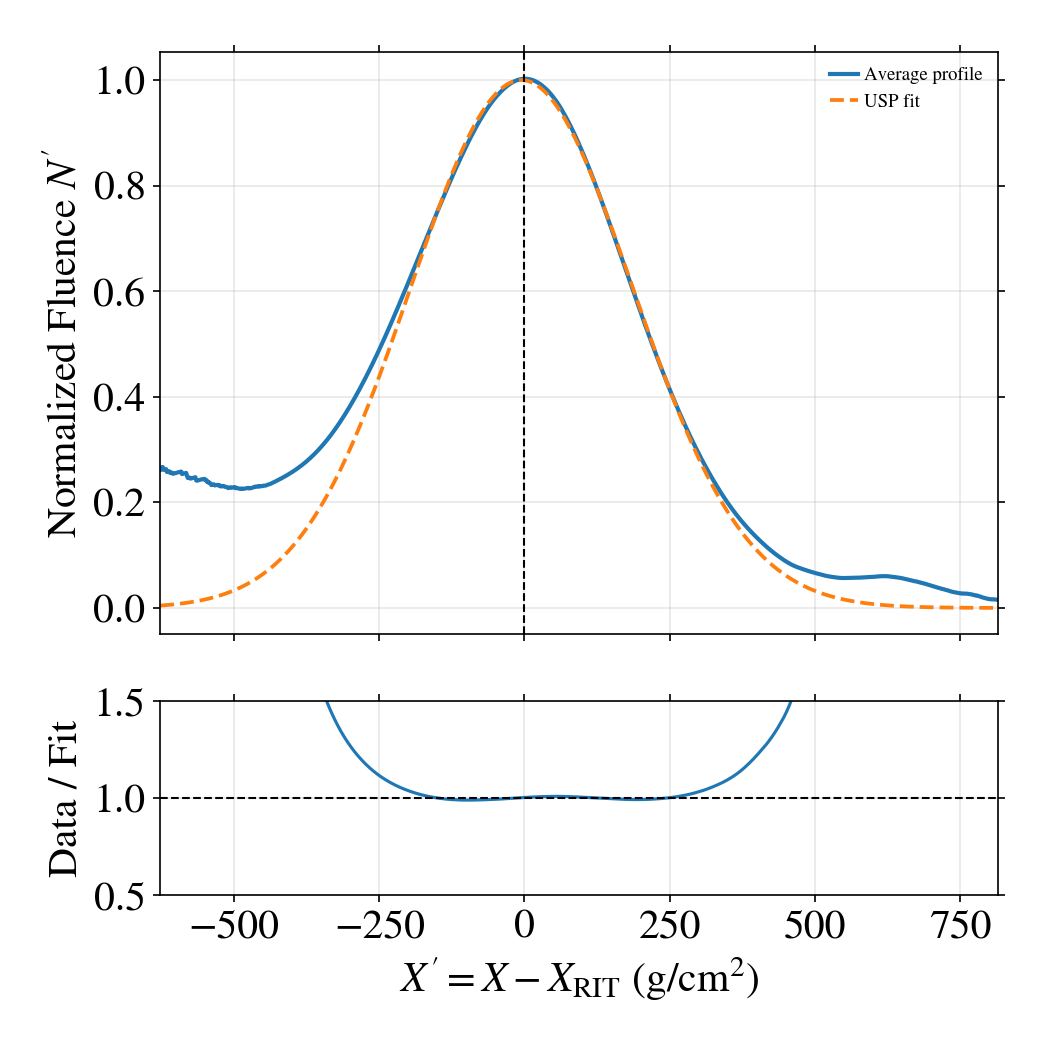}%
 \caption{RIT-derived average profile, applied fit to Eq.\,(\ref{eq:Gaisser-Hillas}), and ratio between them for \textit{SIB Proton} $\theta$ = $50^\circ$.}
 \label{fig:ap_USPvsfitratio_th50}
\end{figure}

In Fig.~\ref{fig:ap_XRIT_vs_Xmax_alldatasets}, we showcase the strong correlation between $\Xrit$ and $\Xmax$ for the four main datasets in this work, all obtained at a zenith angle of $\theta = 30^{\circ}$. Some notable features include the expected more compact distributions within iron datasets when compared to proton ones, as well as their distribution towards lower $\Xmax$ values and the linear correlation between $\Xrit$ and $\Xmax$ distributions, described in detail previously in \cite{RIT1, RIT2}.

\begin{figure}[!h]
 \centering
 \includegraphics[width=0.9\linewidth]{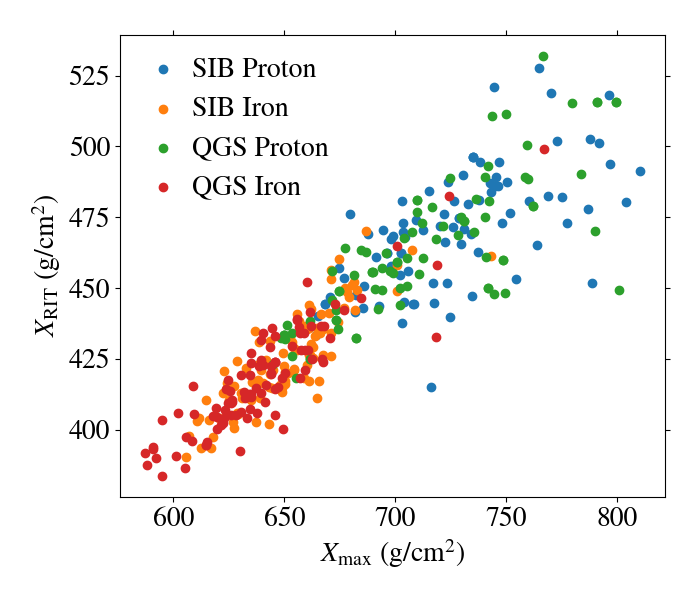}%
 \caption{Correlation between $X_{\rm RIT}$ and $X_{\rm max}$ across the main datasets used in this work, namely \textit{SIB Proton}, \textit{SIB Iron}, \textit{QGS Proton} and \textit{QGS Iron}, all generated with a zenith angle of $30^\circ$.}
 \label{fig:ap_XRIT_vs_Xmax_alldatasets}
\end{figure}

Additionally, in Fig.\,\ref{fig:ap_XRIT_vs_Xmax_diff_zenith}, we show a strong correlation between $\Xrit$ vs $\Xmax$ for the \textit{SIB Proton} datasets simulated with different zenith angles. As seen in \cite{RIT1, RIT2}, higher zenith angle showers will have higher $\Xrit$ values for the same $\Xmax$.

\begin{figure}[!h]
 \centering
 \includegraphics[width=0.9\linewidth]{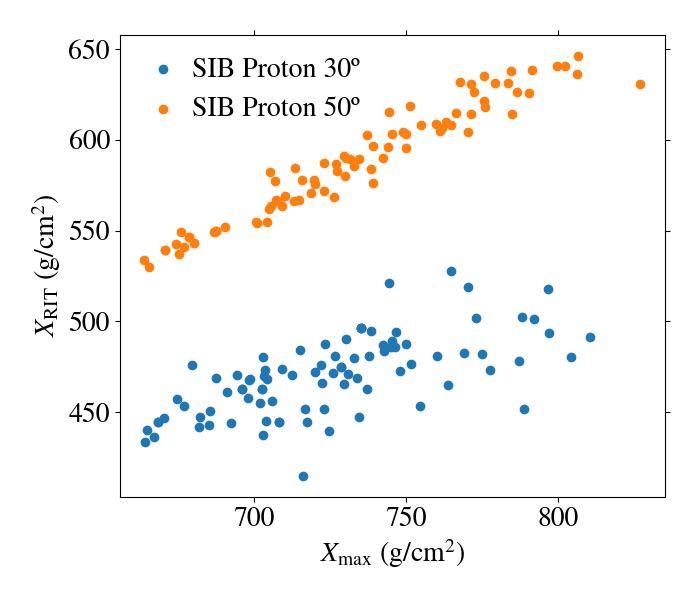}%
 \caption{Correlation between $X_{\rm RIT}$ and $X_{\rm max}$ across two \textit{SIB Proton} datasets, one obtained at a zenith angle of $30^\circ$, and the other at $50^\circ$.}
 \label{fig:ap_XRIT_vs_Xmax_diff_zenith}
\end{figure}

Fig.~\ref{fig:ap_avg_maxdev_map} shows a comprehensive stability study of the fit window. The color scale represents the maximum deviation averaged over datasets between RIT-derived average profiles and their corresponding GH fits. The deviation is evaluated in a variable window defined as [$-50 + \Delta_-$, $50 + \Delta_+$]~g\,cm$^{-2}$. For each dataset, the maximum fit residual within the chosen window is computed and then averaged across datasets. The parameters $\Delta_+$ and $\Delta_-$ extend the base interval $X' \in [-50, 50]$~g\,cm$^{-2}$ separately for $X' > 0$ and $X' < 0$, respectively. This provides a more complete view of the procedure introduced in Fig.\,\ref{fig:window_fits}, clearly delineating the region where the GH description remains stable and the RIT-derived average longitudinal profiles retain a robust GH-like behavior.

\begin{figure}[!h]
 \centering
 \includegraphics[width=0.9\linewidth]{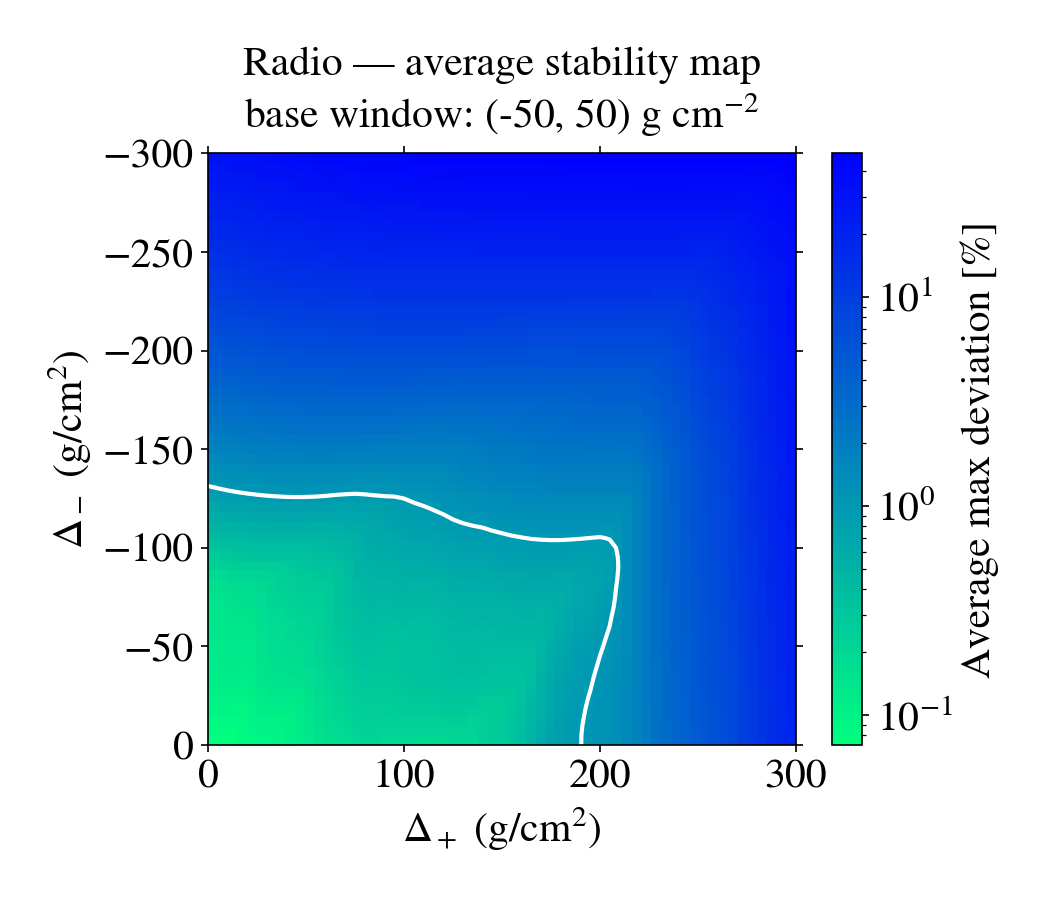}%
 \caption{All main datasets averaged map of the maximum deviation between the RIT-derived average profile and its corresponding fit, obtained by varying the fitting window limits defined as [$-50 + \Delta_-$, $50 + \Delta_+$]~g\,cm$^{-2}$. The white contour line indicates the region where this maximum deviation reaches $1\%$.}
 \label{fig:ap_avg_maxdev_map}
\end{figure}

\end{document}